\documentclass[10pt,aps,prd,floats,floatfix,superscriptaddress,twocolumn,nofootinbib,amsmath,amssymb,showpacs,showkeys]{revtex4-2}
\usepackage[compatibility=false]{caption}
\usepackage[dvipsnames]{xcolor}
\usepackage{multirow}
\usepackage{graphicx}
\usepackage{verbatim}
\usepackage{indentfirst}
\usepackage[normalem]{ulem}
\usepackage{appendix}
\usepackage{soul}
\usepackage{bm}
\usepackage{float}
\usepackage{tikz}
\usepackage{mathtools}
\usepackage{titlesec}
\usepackage{wrapfig}
\usepackage{setspace}
\usepackage{ragged2e}
\usepackage{afterpage}
\usepackage{subcaption}
\usepackage{tablefootnote}
\usepackage{threeparttable}
\usepackage{comment}
\usepackage{booktabs}

\allowdisplaybreaks[2]

\newcommand{\eps}{\varepsilon}
\newcommand{\Nsp}{\mathcal{N}_{\rm sp}}
\newcommand{\Dh}{\bar\Delta^2_h}
\newcommand{\mpl}{m_{\rm pl}}
\newcommand{\Cq}{C_{\rm q}}
\newcommand{\Nobs}{N_{\rm obs}}
\newcommand{\Ntot}{N_{\rm tot}}

\usepackage{hyperref}
\hypersetup{colorlinks=true, linkcolor=black, citecolor=black, urlcolor=black}
\begin{document}

\title{Origin of the red spectral tilt in Inflation without an Inflaton}

\author{Raul Jimenez}
\thanks{raul.jimenez@icc.ub.edu}
\affiliation{ICC, University of Barcelona, Mart\'i i Franqu\`es, 1, E08028 Barcelona, Spain}
\affiliation{ICREA, Pg. Lluis Companys 23, Barcelona, 08010, Spain}

\author{Daniele Bertacca}
\thanks{ daniele.bertacca@pd.infn.it}
\affiliation{Dipartimento di Fisica e Astronomia ``Galileo Galilei'', Universit\`a degli Studi di Padova, via Marzolo 8, I-35131, Padova, Italy}
\affiliation{INFN, Sezione di Padova, via Marzolo 8, I-35131, Padova, Italy}
\affiliation{INAF-Osservatorio Astronomico di Padova, Vicolo dell'Osservatorio 5, I-35122 Padova, Italy}

\author{Mariam Abdelaziz}
\thanks{mariamtarekmohamed.abdelaziz-ssm@unina.it}
\affiliation{Scuola Superiore Meridionale, Largo San Marcellino 10, I-80138 Napoli, Italy}
\affiliation{INFN, Sezione di Napoli, Via Cinthia Edificio 6, I-80126 Napoli, Italy}
\affiliation{Department of Astronomy, Space Science and Meteorology, Cairo University, 12613 Giza, Egypt}

\author{Sabino Matarrese}
\thanks{ sabino.matarrese@pd.infn.it}
\affiliation{Dipartimento di Fisica e Astronomia ``Galileo Galilei'', Universit\`a degli Studi di Padova, via Marzolo 8, I-35131, Padova, Italy}
\affiliation{INFN, Sezione di Padova, via Marzolo 8, I-35131, Padova, Italy}
\affiliation{INAF-Osservatorio Astronomico di Padova, Vicolo dell'Osservatorio 5, I-35122 Padova, Italy}
\affiliation{Gran Sasso Science Institute, Viale F. Crispi 7, I-67100 L'Aquila, Italy}

\author{Angelo Ricciardone}
\thanks{angelo.ricciardone@unipi.it}
\affiliation{Dipartimento di Fisica ``Enrico Fermi'', Universit\`a di Pisa, Pisa I-56127, Italy}
\affiliation{INFN sezione di Pisa, Pisa I-56127, Italy}

\author{Marisol Traforetti}
\thanks{marisol.traforetti@icc.ub.edu}
\affiliation{ICC, University of Barcelona, Mart\'i i Franqu\`es, 1, E08028 Barcelona, Spain}

\begin{abstract}
In Inflation without an Inflaton (IWI), a de Sitter (dS) background drives the inflationary expansion, and the graviton bilinear generates scalar perturbations at second order, with no inflaton or additional fundamental scalar field. 
We investigate how this novel scenario can account for the observed scalar spectral tilt.
The tensor-mode dependence of the scalar power spectrum gives the consistency relation $n_s-1=2n_T$, since it contains four tensor mode functions.
Using the observed scalar spectral tilt $(n_s)_\ast\simeq0.9649$ at Cosmic Microwave Background (CMB) scales, we obtain $n_T\simeq-0.0175$, independent of the tensor-to-scalar ratio $r$, hence different from the standard single-field slow roll prediction $n_T=-r/8$.
A full time-dependent calculation of $\phi_2$ in exact dS leads to an almost scale-invariant scalar power spectrum, with a small blue infrared residual on the order of $n_s-1\simeq+4\times10^{-4}$. 
In order to account for the observed red tilt, we therefore introduce a slow drift in the Hubble rate, parametrized by $\eps=-\dot H/H^2$. The tensor tilt is then given by $n_T=-2\eps$, implying $n_s-1\simeq-4\eps$, at leading order. 
The observed tilt requires $\eps_\ast\simeq0.0088$, far exceeding the drift that can be sustained by the classical graviton backreaction, for which we find $|\eps_{\rm self}|\lesssim10^{-11}$.
Therefore, we investigate the quantum break-time of dS as the origin of the background evolution. In this picture, the scalar tilt becomes $n_s-1=-4\Cq\Nsp(H/\mpl)^2$, where $C_q$ is an order-one depletion coefficient, and the tilt is directly connected to the particle content of dS, $\Nsp$.
The evolution of the drift itself further predicts a positive running of $\alpha_s=+\tfrac12(1-n_s)^2\simeq+6\times10^{-4}$.
These results conclude the distinctive observational predictions of the IWI scenario, establishing it as a falsifiable alternative to standard inflaton-driven models.
\end{abstract}

\maketitle

\section{Introduction}
\label{sec:intro}

The scalar spectral index is the cleanest measured departure of the primordial Universe from exact scale invariance, $n_s=0.9649\pm0.0042$ \cite{Planck:2018vyg,Planck:2018jri}.
In the inflationary paradigm \cite{Starobinsky:1980te,Guth:1980zm,Linde:1981mu,Mukhanov:1981xt}, its value reflects the slow-roll dynamics and the shape of a potential chosen among many that can fit current observations.
The \textit{Inflation without an Inflaton} (IWI) scenario
\cite{Bertacca:2024zfb,Traforetti:2025cax,Abdelaziz:2026qyk}  offers a different picture.
In this case, the inflationary expansion is driven by a
cosmological-constant-like term, without an inflaton or a
scalar potential, while scalar curvature perturbations are
generated at second order by the graviton bilinear. 
Previous works have shown that this mechanism reproduces the
observed scalar amplitude and an adiabatic, nearly Gaussian
spectrum \cite{Bertacca:2024zfb,Traforetti:2025cax}, with non-Gaussian signatures far
below current observational bounds \cite{Abdelaziz:2026qyk}. 
Moreover, the two- and three-point computations of Refs. \cite{Traforetti:2025cax,Abdelaziz:2026qyk} made use of the constant large scale limit solution for $\phi_2$ and the scalar power spectrum resulted in being scale-invariant \cite{Traforetti:2025cax}. 

In this paper, we investigate the origin of the red tilt in the IWI scenario.
In Sec. \ref{sec:setup}, we briefly summarize the perturbative setup. 
In exact de Sitter (dS), dilation symmetry leads to an exactly scale-invariant horizon-crossing spectrum. By extending the calculation beyond the super-horizon approximation, we verify this result in Sec. \ref{sec:Power spectrum in exact de Sitter} using the full time evolution of the scalar perturbations. The fixed comoving infrared cutoff introduces only a small blue residual, $n_s-1\simeq+4\times10^{-4}$. The observed red tilt therefore cannot originate from the time evolution of the scalar perturbations in exact dS and requires a drift in the Hubble rate.
In Sec. \ref{sec:Power spectrum in quasi de Sitter}, we see how a slow drift, parametrized by $\eps=-\dot H/H^2$, introduces a scale dependence in the tensor amplitudes at horizon crossing. Since the scalar power spectrum contains four tensor-mode functions, this gives $n_s-1\simeq-4\eps$, at leading order, i.e., twice the tensor spectral tilt $n_T\simeq-2\eps$. We verify numerically that this scaling is preserved in the full time-dependent calculation.
In Sec. \ref{sec:origin}, we address the origin of the drift $\eps$. First, in Sec. \ref{sec:classical channel}, we discuss whether the required drift can be generated by the graviton population itself, by computing its classical backreaction from the effective energy-momentum tensor. 
We then consider in Sec. \ref{sec:The quantum channel: break-time depletion} the quantum break-time of dS \cite{Dvali:2017eba,Dvali:2013eja} as the origin of the background drift. 
In fact, if one describes dS as a quantum coherent state of gravitons, higher-order
processes of particle production occur due to the graviton's self-coupling and its coupling with other relativistic particle species. This causes the final quantum state to depart from the classical mean-field evolution \cite{Dvali:2017eba} after a timescale $t_q$, the so-called quantum break-time. The intrinsic instability of dS spacetime was studied in previous works \cite{Mottola:1984ar,Antoniadis:2006wq,Polyakov:1982ug,Dvali:2017eba,Mottola:1985ee,Mottola:1985qt,Alicki:2023rfv,Alicki:2023tfz}, and it was argued that it can provide a graceful exit from inflation without the need for a subsequent reheating phase. 
In this framework, the depletion of the dS state transfers energy to a non-vacuum component whose enthalpy drives the decrease of the Hubble rate. Relating the drift to the quantum depletion rate gives $n_s-1\simeq-4\Cq\Nsp \left(\frac{H}{\mpl}\right)^2$, where $\Nsp$ is the number of light gravitationally coupled particle species and $\Cq$ is an order-one depletion coefficient. The measured scalar tilt is thus connected to the particle content of de Sitter and its quantum lifetime.

These results lead to distinct observational predictions, including the consistency relation $n_s-1=2n_T$ between the scalar and tensor spectral tilts, independent of the tensor-to-scalar ratio $r$, and a positive running for a fixed number of species.
Section~\ref{sec:smallscales} discusses the fluid description and the transfer through the end of inflation, while Sec.~\ref{sec:Conclusions} presents the observational implications and conclusions. The technical details of the time-dependent calculation and the graviton effective fluid backreaction are given in the Appendices.

Throughout, $\mpl$ denotes the Planck mass and $m_{\rm pl,red}=\mpl/\sqrt{8\pi}$ the reduced Planck mass.
We use the same tensor normalization as in Refs. \cite{Bertacca:2024zfb,Traforetti:2025cax,Abdelaziz:2026qyk}.

\section{Setup}
\label{sec:setup}

Building on the analysis developed in Refs.~\cite{Bertacca:2024zfb,Traforetti:2025cax,Abdelaziz:2026qyk}, we briefly summarize the ingredients relevant to the present discussion. In the IWI framework, we take the following perturbed metric
\begin{eqnarray}
ds^2 &=& a^2(\eta)\bigg\{ -\left(1 + \psi^{(2)} \right)d\eta^2 \nonumber \\ \quad\quad &&
+\left[\left(1 - \phi^{(2)}\right)\delta_{ij} + \chi^{(1)}_{ij} \right]dx^i dx^j \bigg\}\;,
\end{eqnarray}
where $\chi^{(1)}_{ij}$ are the first-order tensor perturbations, while the scalar fluctuations $\psi^{(2)}$ and $\phi^{(2)}$ arise only at second order. 
Tensor-induced scalars have also been recently studied in Refs. \cite{Bari:2021xvf,Bari:2022grh,Abdelaziz:2025qpn,Abdelaziz:2026sjp}. 
We consider a de Sitter (dS) background and assume Einstein gravity.
The contribution of tensor perturbations is included as an effective fluid on the right-hand side of Einstein's field equations, arising from the vacuum expectation value of the second-order Einstein tensor (see \cite{Bertacca:2024zfb,Traforetti:2025cax,Abdelaziz:2026qyk} for more details).
Expanding Einstein's equations to quadratic order in $\chi^{(1)}_{ij}$, Ref. \cite{Bertacca:2024zfb} obtained the equation of motion for the potential $\phi^{(2)}$. At second order, the scalar mode is sourced by the nonlinear couplings of the first-order tensor modes. For simplicity, in Refs. \cite{Traforetti:2025cax,Abdelaziz:2026qyk}, we used the large-scale limit solution for $\phi_2$ to develop the calculation for the scalar power spectrum $\Delta_\phi^2(k)$ and the bispectrum $B_\phi(k_1,k_2,k_3)$ sourced by tensor perturbations. 
In the present work, we go beyond the super-horizon approximation by including a full time calculation in the evolution of the second-order scalar perturbations. The computation makes use of the Green's function method and the technical details are reported in Appendix \ref{sec: Time-dependent computation}. 

\section{Power spectrum in exact de Sitter}
\label{sec:Power spectrum in exact de Sitter}

In exact dS with constant equation of state $w$ and adiabatic sound speed $c_s$ of the graviton effective fluid, dilations $\eta\rightarrow\lambda\eta$, $\mathbf x\rightarrow\lambda\mathbf x$ are an isometry, and every ingredient of the scalar power spectrum transforms homogeneously. The spectrum, read at per-mode crossing, factorizes as in Eq. \eqref{eq: final expression scalar spectrum}, i.e.
\begin{align}
\Delta^2_\phi\left(k,\eta_\ast(k)\right)&=\mathcal{A}^2_h(k_\ast)\left(\dfrac{k}{k_\ast}\right)^{2n_T}\mathcal{I}(\mu)\,,
\label{eq: schematic expression scalar spectrum}
\end{align}
where $\mathcal{I}(\mu)$ is constructed numerically by performing the loop and time integral, as outlined in Appendix \ref{sec: Time-dependent computation}. In Eq. \ref{eq: schematic expression scalar spectrum}, $\mathcal{A}_h(k_\ast)$ is the tensor amplitude at the pivot, $n_T$ is the tensor tilt, $\mu=k_{\rm min}/k$, and the comoving infrared cutoff $k_{\rm min}$ is the first mode to exit at the onset of inflation. On the other hand, the UV truncation is imposed as a fixed dimensionless loop ratio \footnote{The fixed-ratio support used in the hard-pair realization was introduced in Ref. \cite{Traforetti:2025cax}, where it was interpreted as a non-linear coupling between modes rather than as a cutoff tied to a fixed comoving momentum.} $X_{\rm uv}=k_{\rm max}/k$, hence it does not introduce a scale-dependence in the $k$-sweep. 
The scalar spectral tilt is defined as $$n_s(k)-1\equiv\frac{d\ln\Delta^2_\phi}{d\ln k}\,.$$ Using Eq. \eqref{eq: schematic expression scalar spectrum}, this is therefore given by
\begin{align}
n_s(k)-1=2n_T-\frac{d\ln \mathcal{I}}{d\ln\mu}\bigg|_{\mu=k_{\min}/k}\,.
\label{eq: spectral slope}
\end{align}
However, since the frozen tensor power spectrum carries no spectral tilt in exact dS, $n_T=0$, the only residual scale dependence enters through $\mathcal{I}(\mu)$. We find that the infrared cutoff contribution $\delta_{\rm IR}$ gives only a small blue residual of the order of $4\times10^{-4}$ to the scalar spectral tilt. \footnote{See more details of the computation in Appendix \ref{sec: Time-dependent computation}. }
This completes the constant-$\phi_2$ computation of Ref. \cite{Traforetti:2025cax} and corrects the heuristic argument of Appendix E of \cite{Traforetti:2025cax}, where we raised the possibility that the time-dependence could define the tilt of the scalar power spectrum.
Therefore, we can conclude that whatever produces the observed tilt should be cleanly identifiable. In the next Section, we introduce a background drift $\eps$ in the Hubble parameter and see whether we can account for the observed value of $n_s$ in the IWI scenario.

\section{A drift in the Hubble rate}
\label{sec:Power spectrum in quasi de Sitter}

Since the scalar is a graviton bilinear, the scalar perturbations inherit their scale dependence from the tensor perturbations. 
We saw in Section \ref{sec:Power spectrum in exact de Sitter} that within exact dS the scalar power spectrum is scale-invariant even if we go beyond the large-scale limit of Refs. \cite{Traforetti:2025cax,Abdelaziz:2026qyk}. 
Here, we consider a slow drift in the Hubble rate $H$, parametrized by $\eps=-\dot H/H^2_{\rm inf}>0$, so that $H(\eta)\propto(-\eta)^{\eps}$ to first order in $\eps$. We know that the amplitude of the tensor fluctuations is set by the Hubble rate at horizon crossing. Hence, in exact dS with constant $H$, we obtain an exactly scale-invariant tensor spectrum. On the other hand, if we let $H$ decrease slowly, the Hubble rate at horizon exit of each tensor mode becomes scale-dependent. More precisely, the amplitude of the tensor modes scales as $\propto H(\eta_q)q^{-3/2}\propto q^{-3/2-\eps}$. Consequently, the leading-order tensor tilt is given by $n_T\simeq-2\eps\not=0$.

In the IWI framework, scalar perturbations are generated dynamically at second order by quadratic terms of first-order tensor modes. Their power spectrum therefore contains four tensor modes; hence Eq. \eqref{eq: schematic expression scalar spectrum} becomes
\begin{align}
\Delta^2_\phi\left(k,\eta_\ast(k)\right)&=\mathcal{A}^2_h(k_\ast)\left(\dfrac{k}{k_\ast}\right)^{-4\eps}\mathcal{I}_\eps(\mu)\,.
\label{eq: schematic expression scalar spectrum with epsilon}
\end{align}
In the expression above, the subscript $\eps$ in the kernel $\mathcal{I}_\eps(\mu)$ is meant to emphasize that the remaining leading-order quasi-dS corrections \footnote{To leading order in $\eps$, other modifications include the shift of the tensor Hankel index from $\nu_T=3/2$ to $\nu_T=3/2+\eps$ and the Green's-function corrections.} are included in the rescaled response function, but they do not introduce an additional explicit power of the external wavenumber.
From Eq. \eqref{eq: schematic expression scalar spectrum with epsilon}, it is clear that the spectral tilt is given by
\begin{align}
    n_s-1=2n_T+\dfrac{d\ln\mathcal{I}_\eps}{d\ln k}\simeq-4\eps + \mathcal{O}(\eps^2)+\dfrac{d\ln\mathcal{I}_\eps}{d\ln k}\,.
    \label{eq: spectral tilt with drift}
\end{align}
If the contribution of $\mathcal{I}_\eps(\mu)$ is small, meaning that the dimensionless response associated with the depletion and exit introduces no additional scale dependence over the observable range, then the scale dependence comes primarily from the $-4\eps$ weight from the four tensor legs. In this case, we can write 
\begin{align}
    n_s-1\simeq-4\eps+\mathcal{O}(\eps^2)\,.
     \label{eq: spectral tilt leading order epsilon}
\end{align}
We test numerically that the imposed dilation weight of the tensors is correctly propagated into the scalar power spectrum, given our choices for the reading prescription and UV cutoff entering $\mathcal{I}_\eps(\mu)$. \footnote{see Appendix \ref{sec: Time-dependent computation} for more details}
Moreover, the condition for Eq. \eqref{eq: spectral tilt leading order epsilon} is tested for the exit realization in Ref.~\cite{GWexit}. 
\begin{table}[b]
{
\begin{tabular}{cccc}
\toprule
$\eps$ & $-4\eps$ & $\delta_{\rm IR}$& $n_s-1\simeq-4\eps+\delta_{\rm IR}$  \\
\midrule
$0$ & $0$ & $+4\times10^{-4}$ & $+0.0004$ \\
$0.0050$   & $-0.0200$  & $+5\times10^{-4}$ & $-0.0195$ \\
$0.0088$ & $-0.0351$ & $+5\times10^{-4}$  & $-0.0346$ \\
$0.0089$ & $-0.0356$ & $+5\times10^{-4}$ & $-0.0351$ \\
$0.0200$ & $-0.0800$ & $+6\times10^{-4}$ & $-0.0794$ \\
\bottomrule
\end{tabular}}
\caption{We report the results for the scalar spectral tilt obtained for different values of the drift $\eps$ in the Hubble parameter. For each $\eps$, we show: the leading-order prediction $n_s-1\simeq-4\eps$ (Eq. \ref{eq: spectral tilt leading order epsilon}), the infrared contribution $\delta_{\rm IR}\equiv d\ln\mathcal{I}_\eps(\mu)/d\ln k$, and in the last column the scalar slope including the small blue infrared residual (Eq. \eqref{eq: spectral tilt with drift}). The first row corresponds to the exact dS case discussed in Section \ref{sec:Power spectrum in exact de Sitter}. The value $\eps_\ast\simeq0.0088$ highlights the drift needed to obtain the Planck's spectral tilt, i.e., $(n_s)_\ast=0.9649$.}
\label{tab:spectral tilt results from drift}
\end{table}

\begin{figure*}[t]
\centering
\includegraphics[width=0.94\textwidth]{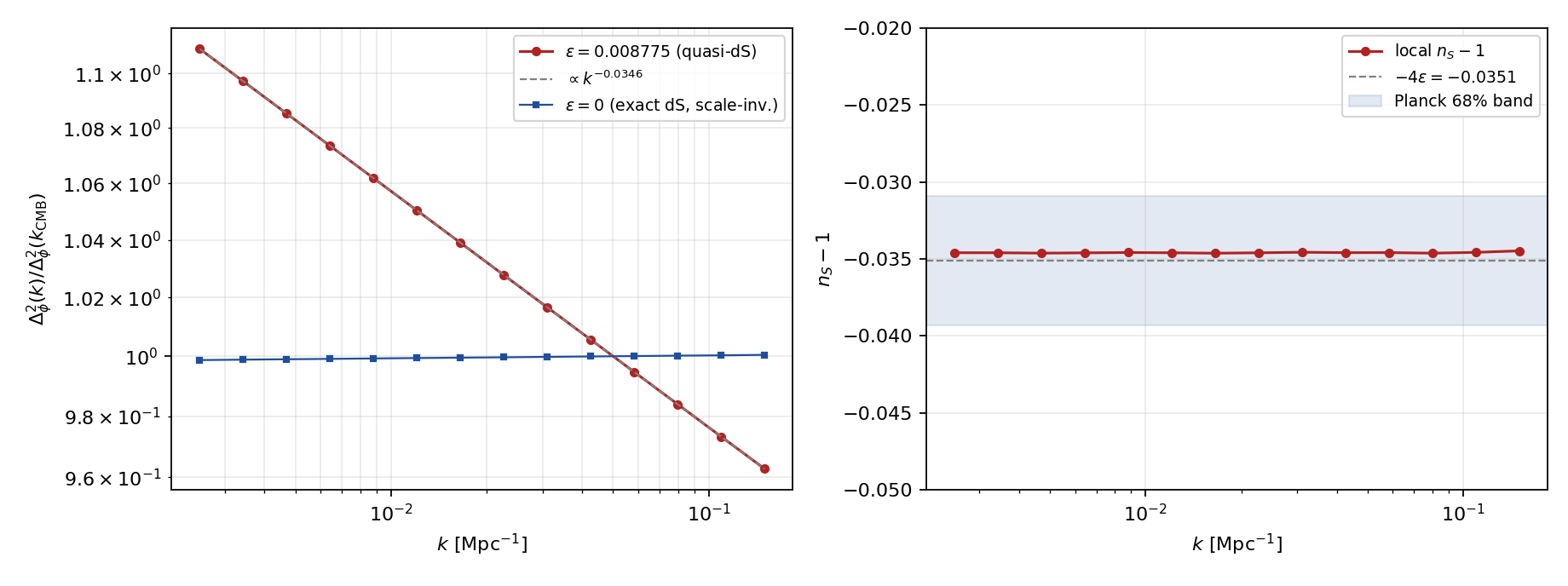}
\caption{Left: Dimensionless scalar power spectrum $\Delta^2_\phi(k)$ as a function of the external scale $k$, and normalized to its value at $k_{\rm CMB}=0.05\,\rm Mpc^{-1}$. The blue line shows the exact dS case ($\eps=0$), which gives a nearly scale-invariant result (see Section \ref{sec:Power spectrum in exact de Sitter}). The red line, instead, shows the quasi-dS case for $\eps_\ast\equiv\eps(k_\ast)\simeq0.0088$, i.e., the value of the drift that reproduces the Planck central tilt $(n_s)_\ast\simeq0.9649$ at the pivot scale $k_\ast=0.01\,\rm Mpc^{-1}$ \cite{Planck:2018jri,Planck:2018vyg}. The dashed line highlights the power-law behavior. Right: The scalar spectral tilt $n_s-1$ as a function of the external wavenumber $k$. The red line shows the numerical result including the small infrared residual (Eq. \eqref{eq: spectral tilt with drift}), which is found to be of the order of $\simeq-0.0346$, constant up to $5\times10^{-5}$ in the tested range of momenta. The dashed line highlights the leading-order prediction, i.e., $n_s-1\simeq-4\eps_\ast=-0.0351$, while the Planck $68\%$ band is shown as a shaded area.}
\label{fig:scalar power spectrum}
\end{figure*}

In Table \ref{tab:spectral tilt results from drift}, we summarize the scalar tilt results for different values of $\eps$. 
In particular, let us take the value of the drift required to reproduce Planck's central value of the tilt $(n_s)_\ast\simeq0.9649$ \cite{Planck:2018jri,Planck:2018vyg} at the CMB pivot scale $k_\ast=0.01\,\rm Mpc^{-1}$, i.e. $(n_s-1)_\ast\simeq-0.0351=-4\eps_\ast$. This corresponds to $\eps_\ast=\eps(k_\ast)\simeq0.0088$, and in this case the neglected higher-order corrections enter at $\mathcal{O}(\eps_\ast^2)\simeq8\times10^{-5}$. Computing the tilt numerically, we find the infrared cutoff residual $d\ln\mathcal{I}_\eps/d\ln k$ is on the order of $+5\times10^{-4}$. Following Eq. \eqref{eq: spectral tilt with drift}, this yields a scalar tilt of $\simeq-0.0346$ with variations of $5\times10^{-5}$ across the tested range of the external momenta $k$. 
We plot the dimensionless power spectrum of scalar perturbations $\Delta^2_\phi(k)$ as a function of the external wave-number $k$ in the left panel of Figure \ref{fig:scalar power spectrum} for a drift $\eps_\ast$ (red line), and we compare it with the exact dS case (blue line). Instead, in the right panel of Figure \ref{fig:scalar power spectrum}, we show the result for the scalar spectral tilt in the reference case $\eps_\ast$.
Finally, for fixed $\eps$, we find no measurable running \footnote{See Eq. \eqref{eq:running} for the expected running induced by a drift $\eps$.} with $|dn_s/d\ln k|\lesssim10^{-4}$, far below Planck's $\alpha_s=-0.0045\pm0.0067$ \cite{Planck:2018jri}. 

\section{The origin of the drift}
\label{sec:origin}
\subsection{The classical channel: graviton backreaction}
\label{sec:classical channel}

Could the tensor modes themselves generate the required background drift?
We have seen in the previous Section that the observed tilt fixes how large this effect must be: the leading-order relation $n_s-1\simeq-4\eps$ requires $\eps_\ast\simeq0.0088\pm0.0011$. 
The tensor modes that generate the scalar fluctuations also backreact on the background expansion. If this backreaction can produce the required decrease in $H$, the scalar perturbations and their red tilt would arise from the same gravitational dynamics. 

We calculate this contribution from the effective energy-momentum tensor of the gravitational waves,
\begin{align}
T^{\mu}{}_{\nu,\rm eff}=-\frac{1}{8\pi G}\left\langle(G^{(2)})^{\mu}{}_{\nu}[\chi,\chi]\right\rangle\,,
\end{align}
evaluated on the unperturbed dS background and including both tensor polarizations. With $x=q|\eta|=q/(aH)$ and $\Dh$ denoting the frozen tensor power per logarithmic interval, the resulting energy density and pressure are
\begin{align}
8\pi G\,\frac{d\rho_{\rm GW}}{d\ln q}&=\frac{\Dh H^2}{8}\,x^2(2x^2-7)\,,\nonumber\\
8\pi G\,\frac{dp_{\rm GW}}{d\ln q}&=\frac{\Dh H^2}{24}\,x^2(2x^2+7)\,.
\label{eq:fluid}
\end{align}
Appendix \ref{sec: The graviton effective fluid} derives these expressions from the tensor mode functions and reproduces them using the effective energy-momentum tensor of Ref. \cite{Abramo:1997hu}.

From Eq. \eqref{eq:fluid}, we see that the $x^4$ terms dominate inside the Hubble radius, and the graviton population behaves as radiation, with $w\to1/3$ \cite{Isaacson:1968hbi}. Outside the Hubble radius, instead, the $x^2$ terms give a negative effective energy density and $w\to-1/3$. Both the density and pressure decay as $x^2$ in this limit. The accumulation of frozen modes therefore gives a finite infrared contribution, with no secular growth in the classical backreaction calculated here.
The behavior of the graviton population is shown in Figure \ref{fig:backreaction} in Appendix \ref{sec: The graviton effective fluid}.

The change in the expansion rate is controlled by the enthalpy, through $\dot H=-(\rho+p)/(2m_{\rm pl,red}^2)$. Since the cosmological constant has $\rho_\Lambda+p_\Lambda=0$, the tensor contribution to the drift $\eps=-\dot H/H^2$ comes directly from $\rho_{\rm GW}+p_{\rm GW}$. Integrating the tensor population over $0<x\leq X$ gives $\eps_{\rm self}(X)=\Dh(2X^4-7X^2)/48$, as derived in Eq.~\eqref{eq:epsselfX}. Here, $X$ sets the upper matching scale near Hubble crossing. For the choice $X=1$, we find
\begin{equation}
\eps_{\rm self}=\frac{(\rho+p)_{\rm GW}}{2m_{\rm pl,red}^2H^2}=-\frac{5}{48}\Dh\,.
\label{eq:epsself}
\end{equation}
Varying the matching scale over $1\leq X\leq2$ changes the sign of the result, but throughout this interval its magnitude remains bounded by
\begin{align}
|\eps_{\rm self}|\lesssim0.13\,\Dh\,.
\end{align}
The tensor amplitude therefore sets the size of the effect.

With the normalization fixed by the scalar amplitude in Ref.~\cite{Traforetti:2025cax}, the tensor power spectrum is $\Dh\simeq3\times10^{-11}$ at $\Nobs=30$ and $\Dh\simeq1.4\times10^{-17}$ at $\Nobs=60$. For $X=1$, this gives $|\eps_{\rm self}|\simeq3.1\times10^{-12}$ and $\simeq1.5\times10^{-18}$, respectively. These values lie nine to sixteen orders of magnitude below the drift required by the observed tilt. Classical graviton backreaction thus leaves pure-$\Lambda$ IWI nearly scale invariant.

The red tilt must therefore originate beyond this classical contribution. Quantum backreaction with secular growth has been studied in Ref.~\cite{Tsamis:1996qq}. In the vacuum-depletion description considered below, the measured tilt tells us how rapidly the vacuum energy must decrease. For vacuum-dominated expansion, $\rho_\Lambda\simeq3m_{\rm pl,red}^2H^2$, so we get
\begin{align}
\frac{d\ln\rho_\Lambda}{dN}\simeq-2\eps\simeq-0.0176\,,
\end{align}
and the required decrease is approximately $1.8\%$ per $e$-fold. The energy released by vacuum depletion contributes to the background density and pressure; we now determine how its enthalpy sustains this evolution.

\subsection{The quantum channel: break-time depletion}
\label{sec:The quantum channel: break-time depletion}

We have seen that the required drift cannot be generated by the classical graviton backreaction computed in \ref{sec:classical channel}. Since there is no additional independent classical source of the drift in the graviton population considered, in this section, we study whether it can be a measurement of the quantum vacuum energy drain.

If one describes dS space as a coherent state of a large number of gravitons \cite{Dvali:2013eja,Dvali:2017eba}, then these gravitons inevitably interact with each other and also couple to other relativistic species. The quantum state is expected to depart from the classical mean field evolution (the classical dS description) after a characteristic time $t_q$, and the energy stored in the inflationary state is transferred into the produced particles. 
Previous works argued that this quantum instability of dS space \cite{Mottola:1984ar,Polyakov:2009nq} provides both a natural way for a graceful exit from inflation and the means to enter a radiation-dominated phase. 
In this framework, the so-called quantum break-time is
\begin{align}
t_q=\dfrac{1}{\Nsp}\dfrac{\mpl^2}{H^3}
\label{eq:tq}
\end{align}
where $\Nsp$ is the number of light gravitationally coupled particle species and $H$ is the energy scale of dS. In Ref.~\cite{Traforetti:2025cax}, we have seen that this time scale imposes an upper bound on the number of particle species in the IWI scenario, as
\begin{align}
    \Nsp<\dfrac{1}{\Nobs}\dfrac{\mpl^2}{H^2}\equiv\mathcal{N}_{\rm sp}^{\rm max}\,,
    \label{eq: upper bound particle species}
\end{align}
where $\Nobs$ denotes the number of observed e-folds.

If we connect the drift $\eps$ to the quantum depletion of dS over a time scale $t_q$, we expect a coherent state of dS to drift at a parametric rate of $\eps\simeq(H)^{-1}/t_q$, so that each $e$-fold consumes a fraction $\eps$ of the dS quantum lifetime. 
Then, Eq. \eqref{eq: spectral tilt leading order epsilon} gives \footnote{The relation neglects the small infrared residual related to $\mathcal{I}_\eps(\mu)$ in Eq. \eqref{eq: spectral tilt with drift} and corrections beyond leading order in $\eps$.}
\begin{align}
    n_s-1\simeq-4C_q\Nsp\left(\dfrac{H}{m_{\rm pl}}\right)^2\,,
    \label{eq: scalar tilt quantum channel}
\end{align}
where we re-introduced explicitly the dependence on $C_q$, which is an order-one depletion coefficient. In this picture, the scalar tilt is therefore directly connected to the particle content of dS, together with the energy scale and the depletion coefficient.

To clarify how the drift $\eps\simeq1/(t_qH)$ is realized in the Einstein equations, notice that a depleting vacuum cannot be conserved by itself. In fact, writing the vacuum energy tensor as $T_v{}^\mu{}_\nu=-\rho_v\delta^\mu{}_\nu$, one has $\nabla_\mu T_v{}^\mu{}_\nu=-\partial_\nu\rho_v$. Thus, the Bianchi identity requires a non-vacuum sector satisfying $\nabla_\mu T_e{}^\mu{}_\nu=+\partial_\nu\rho_v$, to be consistent with covariant energy conservation, and its enthalpy realizes the background drift, $\eps=\frac{3}{2}(\rho_e+p_e)/\rho$ \cite{GWexit}. 
The effective stress-energy of the graviton population $T^\mu{}_{\nu,\mathrm{eff}}\propto-\langle G^{(2)\mu}{}_\nu[\chi,\chi]\rangle$ should not be confused with the non-vacuum sector produced by vacuum depletion, which need not consist only of gravitons and whose microscopic composition is part of the depletion dynamics. In particular, $T^\mu{}_{\nu,\mathrm{eff}}$ is the stress-energy tensor carried by the graviton population, and it is too small to generate the required drift (see Section \ref{sec:classical channel}); while $T_e{}^\mu{}_\nu$ is required by energy-momentum conservation once $\rho_v$ changes and it carries the background enthalpy associated with $\eps$. 

Then, similarly to the species bound in Eq. \eqref{eq: upper bound particle species}, we can use $\eps\simeq1/(t_qH)=C_q\Nsp(H/m_{\rm pl})^2$ to obtain a bound for the allowed scale of the tilt. From $N_{\rm tot}\leq t_qH$ and $N_{\rm tot}\geq N_{\rm obs}$, one obtains
\begin{align}
    1-n_s\simeq4\eps\leq\dfrac{4C_q}{N_{\rm obs}}\,.
    \label{eq: upper bound tilt}
\end{align}
Thus, the quantum-lifetime bound fixes the allowed scale of the scalar tilt, and a percent-level value \footnote{For example, if we take $C_q=1$ and $N_{\rm obs}=60$, from Eq. \eqref{eq: upper bound tilt} we get $1-n_s\leq0.07$.} of $1-n_s$ follows only if inflation lasts an order-one fraction of $t_q$. 
In particular, treating $\eps$ as constant, saturation
of the quantum lifetime gives $\eps=\Cq/\Ntot$, and therefore
\begin{align}
    1-n_s\simeq4\frac{\Cq}{\Ntot}\,,
    \label{eq: saturation case}
\end{align}
where $\Ntot$ is the total number of $e$-folds \footnote{
When the evolution of the drift itself, $d\eps/dN=-2\eps^2$, is included,
with $N=\ln a$, saturation must instead be imposed
cumulatively as $\int_0^{\Ntot}\eps\,dN=\Cq$.
Taking $\eps_\ast$ as the drift at the CMB pivot
and $N_\ast$ as the number of $e$-folds between
the pivot scale and the end of inflation, we obtain
$\Ntot=(1/(2\eps_\ast)+N_\ast)(1-e^{-2\Cq})$.}.

Now let us check the implications of this picture on the allowed duration of inflation. 
For the Planck central value $(n_s)_\ast=0.9649$ and $\Cq=1$ in the constant-$\eps$ approximation, saturation in \eqref{eq: saturation case} corresponds to $\Ntot\simeq114$ $e$-folds.
This can be compared to the maximum duration of inflation of $\simeq126$ $e$-folds obtained in \cite{Traforetti:2025cax} using laboratory constraints on the number of species \cite{Dvali:2007hz,Dvali:2007wp}.

\begin{figure}[t]
\centering\includegraphics[width=\columnwidth]{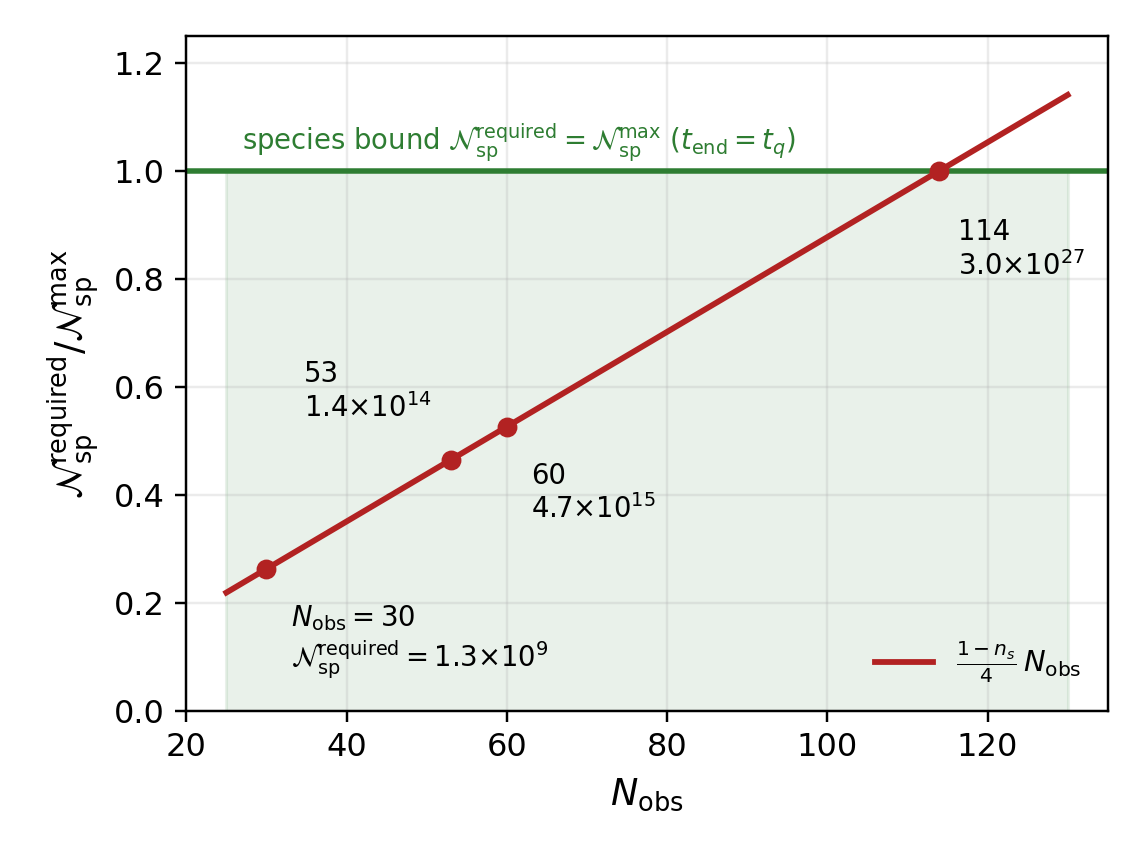}
\caption{We plot the number of species required by the tilt, $\mathcal{N}_{\rm sp}^{\rm required}$ (Eq. \ref{eq: required species count}), in units of the maximum number of species $\mathcal{N}_{\rm sp}^{\rm max}$ (Eq. \ref{eq: upper bound particle species}) obtained in Ref. \cite{Traforetti:2025cax}, as a function of the number of observed $e$-folds. It is clear that following Eq. \ref{eq: ratio particle species}, the ratio lies below unity until saturation is reached. Numerical results of the points highlighted for $N_{\rm obs}=30,53,60,114$ can be compared with Table \ref{tab: numerical results}. The green shaded area highlights the region $N_{\rm obs}\in(30,126)$ from Ref. \cite{Traforetti:2025cax}.}
\label{fig:particle species ratio VS Nobs}
\end{figure}

Then, from Eq. \eqref{eq: scalar tilt quantum channel} the number of species required by the observed tilt can be written as
\begin{align}
\Nsp^{\rm required}=\frac{1-n_s}{4}\left(\frac{m_{\rm pl}}{H}\right)^2\,.
\label{eq: required species count}
\end{align}
We can check that $\mathcal{N}_{\rm sp}^{\rm required}$ lies within the IWI's own bound $\mathcal{N}_{\rm sp}^{\rm max}$, i.e., Eq. \eqref{eq: upper bound particle species}. More precisely, assuming $C_q=1$, the two are related by 
\begin{align}
    \dfrac{\Nsp^{\rm required}}{\Nsp^{\rm max}}=\dfrac{1-n_s}{4}N_{\rm obs}\,,
    \label{eq: ratio particle species}
\end{align}
The result for the ratio $\Nsp^{\rm required}/\Nsp^{\rm max}$ as a function of the number of observed $e$-folds is given in Figure \ref{fig:particle species ratio VS Nobs}. The numerical results for $\Nsp^{\rm required}$ and $\Nsp^{\rm max}$ at a given $N_{\rm obs}$ are reported in Table \ref{tab: numerical results}. In particular, for $N_{\rm obs}=30-60$ and $(n_s)_\ast=0.9649$, we get $\Nsp^{\rm required}/\Nsp^{\rm max}=0.26-0.53$. 

\begin{table*}[t]
\caption{For different values of the number of observed $e$-folds $\Nobs$, we report the numerical results for: the inflationary energy scale $H$, the tensor power spectrum $\Delta^2_h$, the tensor-to-scalar ratio $r$, the number of species required to match Planck's central value $(n_s)_\ast=0.9649$, i.e. $\Nsp^{\rm required}$ (Eq. \eqref{eq: required species count}), and maximum number of allowed species $\Nsp^{\rm max}$ (Eq. \ref{eq: upper bound particle species}, Ref. \cite{Traforetti:2025cax}). To compute $\Nsp^{\rm required}$, we took for simplicity $C_q=1$. In particular, $\Nobs=53$ is the instant reheating case, while $\Nobs=114$ is the saturation case (see Eq. \ref{eq: saturation case}).}
\label{tab: numerical results}
\begin{tabular}{lccccc}
\toprule
$\Nobs$ & $H\,[\rm GeV]$ & $\Delta^2_h$ & $r$ & $\Nsp^{\rm required}$ &
$\Nsp^{\max}$ \cite{Traforetti:2025cax}\\
\midrule
$30$ & $3.2\times10^{13}$ & $3.4\times10^{-11}$ & $1.6\times10^{-2}$ &
$1.3\times10^{9}$ & $5\times10^{9}$\\
$53$ & $9.7\times10^{10}$ & $3.2\times10^{-16}$ &
$1.5\times10^{-7}$ & $1.4\times10^{14}$ & $3\times10^{14}$\\
$60$ & $1.7\times10^{10}$ & $9.4\times10^{-18}$ & $4.5\times10^{-9}$ &
$4.7\times10^{15}$ & $9\times10^{15}$\\
$114$ & $2.1\times10^{4}$ & $1.5\times10^{-29}$ &
$7.0\times10^{-21}$ & $3.0\times10^{27}$ & $3\times10^{27}$\\
\bottomrule
\end{tabular}
\end{table*}

Finally, notice that, given a fixed $\Nsp$, the assumed relation $\eps\propto H^2$ makes $\eps$ decrease as the Hubble rate decreases. Hence, unlike standard slow-roll models, this drift does not evolve towards unity, and inflation ends through the quantum-break transition of the condensate\footnote{This will be addressed in more detail in \cite{GWexit}.}. The same decrease in $\eps$ is what provides a positive running $\alpha_s$. At leading order, it is straightforward to show that we obtain
\begin{align}
    \alpha_s\equiv\frac{dn_s}{d\ln k}\simeq\frac12(1-n_s)^2\,,
\label{eq:running}
\end{align}
So once $n_s$ is measured, the running is fixed.
For $(n_s)_\ast=0.9649$, this gives $\alpha_s\simeq+6.2\times10^{-4}$, well within the uncertainty reported by Planck, $\alpha_s=-0.0045\pm0.0067$ \cite{Planck:2018jri}.

Moreover, it is worth mentioning that in this Section we took the depletion coefficient $C_q$ to be of order one for our numerical results, since it has not yet been derived microscopically. A state-resolved description of the depletion and of its memory is devoted to future work \cite{bridge}, and its quantitative connection to the finite window remains to be established. In Ref. \cite{bridge}, an independent derivation of $1-n_s$ also connects the outlined mechanism to quantum Fisher cosmology \cite{Gomez:2020rqv,Gomez:2021jcl}.

\section{Fluid response and the exit from de Sitter}
\label{sec:smallscales}

Within the restricted fluid closure\footnote{This should not be interpreted as a general property of the depletion-produced sector, whose sound speed, entropy perturbation, anisotropic stress, and possible memory response are microscopic properties to be derived.} adopted here, in which $\Pi_2=\delta^2S=0$ and $c_s^2=1$, the horizon-crossing reading is the self-consistent prescription for the tilt. Here, recall that $\Pi_2$ is the second-order scalar part of the anisotropic stress, $\delta^2S$ is the entropic perturbation, and $c_s^2$ is the adiabatic sound speed squared. 
A more general second-order fluid response, including anisotropic stress and non-adiabatic pressure, can be found in the framework of Ref. \cite{Bertacca:2024zfb}. 
In the horizon-crossing reading prescription, the tensor pairs have internal momenta of the order of the external scale, while the internal momenta that are still sub-horizon at each horizon crossing add vacuum dressing that dominates the normalization of $\mathcal{I}(k)$, but do not contribute to the scale-dependence of the spectrum. 
Moreover, the non-vacuum sector required by the depletion can also take part in the finite-range response that realizes the window; whether it does so remains an open microscopic question in Ref. \cite{GWexit}. If it contributes, any resulting scale dependence will enter as part of $\mathcal I(k)$.

Connecting to the small-scale picture, it is also worth mentioning that the effective graviton gas should be distinguished from this depletion-produced sector. In exact dS, it behaves as a steady-state population\footnote{This can be connected to the condition $w-c_s^2>0$ discussed in Ref. \cite{Bertacca:2024zfb} required for the scalar-growth mechanism.}, in which individual contributions evolve while new modes keep crossing the horizon.
The background is renormalized by the long-wavelength modulation of the graviton-gas energy density and the graviton-pair decoherence \cite{Ning:2023ybc,bridge} motivates the background renormalization to be a purely classical computation.

Finally, the scalar potential $\phi_2$ should be explicitly mapped at crossing to the late-time curvature perturbation $\zeta$. Writing
\begin{equation}
\Delta_\zeta^2(k)=
|T_{\zeta\phi}(k)|^2\Delta_\phi^2(k)\,,
\end{equation}
For the exit realization studied in Ref. \cite{GWexit}, the transfer from $\phi_2$ to the post-exit uniform-density curvature is found to be scale-independent over the range tested and therefore does not modify the tilt.

\section{Conclusions}
\label{sec:Conclusions}

In this work, we have investigated the origin of the red scalar spectral tilt in the Inflation without an Inflaton (IWI) scenario. By extending the calculation of the scalar power spectrum beyond the super-horizon approximation used in Refs. \cite{Traforetti:2025cax,Abdelaziz:2026qyk}, we find that dilation symmetry leads to an exactly scale-invariant spectrum in de Sitter (dS). The fixed comoving infrared cutoff introduces only a small blue residual, which cannot account for the observed red tilt. Reproducing the observed spectral index therefore requires a slow drift $\eps$ in the Hubble rate $H$. Since the scalar perturbations arise from the graviton bilinear, their power spectrum contains four tensor-mode functions, giving $n_s-1\simeq-4\eps$ at leading order. Hence, the red tilt is directly connected to the evolution of the inflationary background.

We have also examined whether the graviton population itself can generate the required drift through its classical backreaction. Our calculation of the graviton effective stress-energy tensor shows that this contribution is nine to sixteen orders of magnitude below the required drift $\eps_\ast\simeq0.0088$. We then considered the quantum break-time of dS as the origin of the background evolution. In this description, the depletion of the dS state transfers energy to a non-vacuum component whose enthalpy drives the decrease of $H$. The resulting relation $n_s-1\simeq-4\Cq\Nsp(H/\mpl)^2$ connects the measured scalar tilt to the number of light particle species $\Nsp$, the inflationary energy scale $H$, and the depletion coefficient $C_q$. With $H$ determined by the scalar amplitude, the observed tilt fixes the required species count for a given $ C_q$.

The quantum lifetime of dS also constrains the duration of inflation and the allowed scale of the tilt. Requiring inflation to last throughout the observable window gives the upper bound in Eq.~\eqref{eq: upper bound tilt}. At saturation, the constant-$\eps$ approximation gives $1-n_s\simeq4\Cq/\Ntot$, corresponding to $\Ntot\simeq114$ for the Planck central value and $\Cq=1$. For the benchmark durations considered, the number of species required to reproduce the observed tilt remains within the quantum-lifetime bound.

The same background drift determines the spectral tilts of both scalar and tensor perturbations. For a window fixed in $e$-folds, with a scale-independent transfer through the end of inflation, we obtain the leading-order consistency relation
\begin{equation}
    n_s-1\simeq2n_T\,.
    \label{eq:consistency}
\end{equation}
For the scalar spectral index $n_s=0.9649\pm0.0042$ \cite{Planck:2018vyg,Planck:2018jri}, the tensor tilt then results in $n_T\simeq-0.0175\pm0.0021$.
Unlike the single-field slow-roll relation $n_T\simeq-r/8$, the tensor tilt in IWI is determined by the scalar tilt, independently of $r$. For fixed $\Nsp$ and $\Cq$, the evolution of $\eps$ also predicts a positive running, $\alpha_s\simeq+6.2\times10^{-4}$. 
Moreover, in the IWI framework, the primordial non-Gaussianity remains strongly suppressed, with $|f_{\rm NL}|\lesssim10^{-14}$ \cite{Abdelaziz:2026qyk,GWexit}. These predictions provide distinct tests of the scenario: a tensor tilt or running inconsistent with the predicted relations, or primordial non-Gaussianity at $|f_{\rm NL}|=\mathcal O(1)$, would rule out this realization.

Finally, a microscopic description of the quantum depletion of dS is the next step in establishing the full evolution of the inflationary state. In particular, the depletion coefficient $\Cq$ and the effective stress-energy of the produced particles remain to be determined, together with their possible contribution to the scalar response and the ultraviolet support of the power-spectrum amplitude \cite{bridge,GWexit}.
These calculations would establish how the quantum evolution of the dS state determines both the background drift responsible for the red tilt and the normalization of the primordial scalar spectrum.

\begin{acknowledgments}
The work of RJ is supported by the Simons Foundation. Funding for the work of MT and RJ was partially provided by project PID2022-141125NB-I00, and grant CEX2024-001451-M funded by MICIU/AEI/10.13039/501100011033. DB and SM acknowledge support from the COSMOS network (www.cosmosnet.it) through ASI (Italian Space Agency) Grants 2016-24-H.0, 2016-24-H.1-2018 and 2020-9-HH.0. 
This work is supported in part by the MUR Departments of Excellence grant “Quantum Frontiers” of the Physics and Astronomy Department of Padova, Italy.
MA acknowledges the hospitality of the Institute of Cosmos Sciences of the University of Barcelona (ICCUB), where this work was carried out. The project that gave rise to these results received the support of a fellowship from ``la Caixa'' Foundation (ID 100010434), awarded to MT with code LCF/BQ/DI24/12070012. \\

AI usage: Claude Code was used to assist with coding, computations, and graphic generation. The ideas, inferences, and results all originate in the (human) brains of the authors. Claude and ChatGPT were used to check citations, spelling, grammar, and inconsistencies. The text was written by the authors, who are responsible for the results, findings, and conclusions of this article.
\end{acknowledgments}

\bibliographystyle{apsrev4-2}
\bibliography{references}

@article{bridge,
    author = "Bertacca, Daniele and Jimenez, Raul and Matarrese, Sabino",
    title = "{Quantum Fisher information, graviton-pair decoherence, and the scalar spectral index in inflation without an inflaton}",
    journal = {in preparation},
    year = 2026
}

@article{GWexit,
    author = "Bertacca, Daniele and Jimenez, Raul and Matarrese, Sabino",
    title = "{Gravitational waves and the exit from de Sitter in inflation without an inflaton}",
    journal = {in preparation},
    year = 2026
}

@article{Planck:2018jri,
    author = "Akrami, Y. and others",
    collaboration = "Planck",
    title = "{Planck 2018 results. X. Constraints on inflation}",
    eprint = "1807.06211",
    archivePrefix = "arXiv",
    primaryClass = "astro-ph.CO",
    doi = "10.1051/0004-6361/201833887",
    journal = "Astron. Astrophys.",
    volume = "641",
    pages = "A10",
    year = "2020"
}

@article{Planck:2018vyg,
    author = "Aghanim, N. and others",
    collaboration = "Planck",
    title = "{Planck 2018 results. VI. Cosmological parameters}",
    eprint = "1807.06209",
    archivePrefix = "arXiv",
    primaryClass = "astro-ph.CO",
    doi = "10.1051/0004-6361/201833910",
    journal = "Astron. Astrophys.",
    volume = "641",
    pages = "A6",
    year = "2020",
    note = "[Erratum: Astron.Astrophys. 652, C4 (2021)]"
}

@article{Starobinsky:1980te,
    author = "Starobinsky, Alexei A.",
    editor = "Khalatnikov, I. M. and Mineev, V. P.",
    title = "{A New Type of Isotropic Cosmological Models Without Singularity}",
    doi = "10.1016/0370-2693(80)90670-X",
    journal = "Phys. Lett. B",
    volume = "91",
    pages = "99--102",
    year = "1980"
}

@article{Mukhanov:1981xt,
    author = "Mukhanov, Viatcheslav F. and Chibisov, G. V.",
    title = "{Quantum Fluctuations and a Nonsingular Universe}",
    journal = "JETP Lett.",
    volume = "33",
    pages = "532--535",
    year = "1981"
}

@article{Guth:1980zm,
    author = "Guth, Alan H.",
    editor = "Fang, Li-Zhi and Ruffini, R.",
    title = "{The Inflationary Universe: A Possible Solution to the Horizon and Flatness Problems}",
    reportNumber = "SLAC-PUB-2576",
    doi = "10.1103/PhysRevD.23.347",
    journal = "Phys. Rev. D",
    volume = "23",
    pages = "347--356",
    year = "1981"
}

@article{Linde:1981mu,
    author = "Linde, Andrei D.",
    editor = "Fang, Li-Zhi and Ruffini, R.",
    title = "{A New Inflationary Universe Scenario: A Possible Solution of the Horizon, Flatness, Homogeneity, Isotropy and Primordial Monopole Problems}",
    reportNumber = "LEBEDEV-81-229",
    doi = "10.1016/0370-2693(82)91219-9",
    journal = "Phys. Lett. B",
    volume = "108",
    pages = "389--393",
    year = "1982"
}

@article{Gomez:2020rqv,
    author = "G{\'o}mez, C{\'e}sar and Jimenez, Raul",
    title = "{The quantum cosmological tilt and the origin of dark matter}",
    eprint = "2012.07883",
    archivePrefix = "arXiv",
    primaryClass = "hep-th",
    doi = "10.1088/1475-7516/2021/10/055",
    journal = "JCAP",
    volume = "10",
    pages = "055",
    year = "2021"
}

@article{Gomez:2021jcl,
    author = "Gomez, Cesar and Jimenez, Raul",
    title = "{The quantum de Sitter root of quasi de Sitter observables}",
    eprint = "2111.05380",
    archivePrefix = "arXiv",
    primaryClass = "hep-th",
    doi = "10.1016/j.dark.2022.101035",
    journal = "Phys. Dark Univ.",
    volume = "36",
    pages = "101035",
    year = "2022"
}

@article{Bari:2021xvf,
    author = "Bari, Pritha and Ricciardone, Angelo and Bartolo, Nicola and Bertacca, Daniele and Matarrese, Sabino",
    title = "{Signatures of Primordial Gravitational Waves on the Large-Scale Structure of the Universe}",
    eprint = "2111.06884",
    archivePrefix = "arXiv",
    primaryClass = "astro-ph.CO",
    doi = "10.1103/PhysRevLett.129.091301",
    journal = "Phys. Rev. Lett.",
    volume = "129",
    number = "9",
    pages = "091301",
    year = "2022"
}

@article{Bari:2022grh,
    author = "Bari, Pritha and Bertacca, Daniele and Bartolo, Nicola and Ricciardone, Angelo and Giardiello, Serena and Matarrese, Sabino",
    title = "{An analytical study of the primordial gravitational-wave-induced contribution to the large-scale structure of the Universe}",
    eprint = "2209.05329",
    archivePrefix = "arXiv",
    primaryClass = "astro-ph.CO",
    doi = "10.1088/1475-7516/2023/07/034",
    journal = "JCAP",
    volume = "07",
    pages = "034",
    year = "2023"
}

@article{Mottola:1984ar,
    author = "Mottola, E.",
    title = "{Particle Creation in de Sitter Space}",
    reportNumber = "NSF-ITP-84-123",
    doi = "10.1103/PhysRevD.31.754",
    journal = "Phys. Rev. D",
    volume = "31",
    pages = "754",
    year = "1985"
}

@article{Antoniadis:2006wq,
    author = "Antoniadis, Ignatios and Mazur, Pawel O. and Mottola, Emil",
    title = "{Cosmological dark energy: Prospects for a dynamical theory}",
    eprint = "gr-qc/0612068",
    archivePrefix = "arXiv",
    reportNumber = "LA-UR-06-8631, CERN-PH-TH-2006-251",
    doi = "10.1088/1367-2630/9/1/011",
    journal = "New J. Phys.",
    volume = "9",
    pages = "11",
    year = "2007"
}

@article{Polyakov:1982ug,
    author = "Polyakov, Alexander m.",
    title = "{PHASE TRANSITIONS AND THE UNIVERSE}",
    doi = "10.1070/PU1982v025n03ABEH004529",
    journal = "Sov. Phys. Usp.",
    volume = "25",
    pages = "187",
    year = "1982"
}

@article{Polyakov:2009nq,
    author = "Polyakov, A. M.",
    title = "{Decay of Vacuum Energy}",
    eprint = "0912.5503",
    archivePrefix = "arXiv",
    primaryClass = "hep-th",
    reportNumber = "PUPT-2320",
    doi = "10.1016/j.nuclphysb.2010.03.021",
    journal = "Nucl. Phys. B",
    volume = "834",
    pages = "316--329",
    year = "2010"
}

@article{Dvali:2013eja,
    author = "Dvali, Gia and Gomez, Cesar",
    title = "{Quantum Compositeness of Gravity: Black Holes, AdS and Inflation}",
    eprint = "1312.4795",
    archivePrefix = "arXiv",
    primaryClass = "hep-th",
    doi = "10.1088/1475-7516/2014/01/023",
    journal = "JCAP",
    volume = "01",
    pages = "023",
    year = "2014"
}

@article{Dvali:2017eba,
    author = "Dvali, Gia and Gomez, Cesar and Zell, Sebastian",
    title = "{Quantum Break-Time of de Sitter}",
    eprint = "1701.08776",
    archivePrefix = "arXiv",
    primaryClass = "hep-th",
    reportNumber = "LMU-ASC-08-17, MPP-2017-10",
    doi = "10.1088/1475-7516/2017/06/028",
    journal = "JCAP",
    volume = "06",
    pages = "028",
    year = "2017"
}

@article{Mottola:1985ee,
    author = "Mottola, Emil",
    title = "{A Quantum Fluctuation Dissipation Theorem for General Relativity}",
    reportNumber = "NSF-ITP-85-35",
    doi = "10.1103/PhysRevD.33.2136",
    journal = "Phys. Rev. D",
    volume = "33",
    pages = "2136",
    year = "1986"
}

@article{Mottola:1985qt,
    author = "Mottola, E.",
    title = "{THERMODYNAMIC INSTABILITY OF DE SITTER SPACE}",
    doi = "10.1103/PhysRevD.33.1616",
    journal = "Phys. Rev. D",
    volume = "33",
    pages = "1616--1621",
    year = "1986"
}

@article{Alicki:2023rfv,
    author = "Alicki, Robert and Barenboim, Gabriela and Jenkins, Alejandro",
    title = "{Quantum thermodynamics of de Sitter space}",
    eprint = "2307.04800",
    archivePrefix = "arXiv",
    primaryClass = "gr-qc",
    reportNumber = "IFIC/23-31",
    doi = "10.1103/PhysRevD.108.123530",
    journal = "Phys. Rev. D",
    volume = "108",
    number = "12",
    pages = "123530",
    year = "2023"
}

@article{Alicki:2023tfz,
    author = "Alicki, Robert and Barenboim, Gabriela and Jenkins, Alejandro",
    title = "{The irreversible relaxation of inflation}",
    eprint = "2307.04803",
    archivePrefix = "arXiv",
    primaryClass = "gr-qc",
    reportNumber = "IFIC/23-29",
    doi = "10.1016/j.physletb.2025.139519",
    journal = "Phys. Lett. B",
    volume = "866",
    pages = "139519",
    year = "2025"
}

@article{Dvali:2007hz,
    author = "Dvali, Gia",
    title = "{Black Holes and Large N Species Solution to the Hierarchy Problem}",
    eprint = "0706.2050",
    archivePrefix = "arXiv",
    primaryClass = "hep-th",
    doi = "10.1002/prop.201000009",
    journal = "Fortsch. Phys.",
    volume = "58",
    pages = "528--536",
    year = "2010"
}

@article{Dvali:2007wp,
    author = "Dvali, Gia and Redi, Michele",
    title = "{Black Hole Bound on the Number of Species and Quantum Gravity at LHC}",
    eprint = "0710.4344",
    archivePrefix = "arXiv",
    primaryClass = "hep-th",
    doi = "10.1103/PhysRevD.77.045027",
    journal = "Phys. Rev. D",
    volume = "77",
    pages = "045027",
    year = "2008"
}

@article{Abdelaziz:2025qpn,
    author = "Abdelaziz, Mariam and Bari, Pritha and Matarrese, Sabino and Ricciardone, Angelo",
    title = "{Non-Gaussianity of tensor induced density perturbations}",
    eprint = "2504.07063",
    archivePrefix = "arXiv",
    primaryClass = "astro-ph.CO",
    doi = "10.1103/bb22-pq2m",
    journal = "Phys. Rev. D",
    volume = "112",
    number = "2",
    pages = "023505",
    year = "2025"
}

@article{Abdelaziz:2026sjp,
    author = "Abdelaziz, Mariam and Bari, Pritha and Matarrese, Sabino and Ricciardone, Angelo",
    title = "{Tracing Primordial Gravitational Waves via non-Gaussian Signatures of Halo Bias}",
    eprint = "2605.02882",
    archivePrefix = "arXiv",
    primaryClass = "astro-ph.CO",
    journal=" ",
    month = "5",
    year = "2026"
}

@article{Bertacca:2024zfb,
    author = "Bertacca, Daniele and Jimenez, Raul and Matarrese, Sabino and Ricciardone, Angelo",
    title = "{Inflation without an inflaton}",
    eprint = "2412.14265",
    archivePrefix = "arXiv",
    primaryClass = "astro-ph.CO",
    doi = "10.1103/vfny-pgc2",
    journal = "Phys. Rev. Res.",
    volume = "7",
    number = "3",
    pages = "L032010",
    year = "2025"
}

@article{Traforetti:2025cax,
    author = "Traforetti, Marisol and Abdelaziz, Mariam and Bertacca, Daniele and Jimenez, Raul and Matarrese, Sabino and Ricciardone, Angelo",
    title = "{Inflation without an inflaton. II. Observational predictions}",
    eprint = "2511.11808",
    archivePrefix = "arXiv",
    primaryClass = "astro-ph.CO",
    doi = "10.1103/cfg4-5p3s",
    journal = "Phys. Rev. D",
    volume = "113",
    number = "2",
    pages = "023553",
    year = "2026"
}

@article{Abdelaziz:2026qyk,
    author = "Abdelaziz, Mariam and Traforetti, Marisol and Bertacca, Daniele and Jimenez, Raul and Matarrese, Sabino and Ricciardone, Angelo",
    title = "{Inflation without an inflaton. III. Non-Gaussian signatures}",
    eprint = "2603.15036",
    archivePrefix = "arXiv",
    primaryClass = "astro-ph.CO",
    doi = "10.1103/v7dk-q18l",
    journal = "Phys. Rev. D",
    volume = "114",
    number = "2",
    pages = "023517",
    year = "2026"
}

@article{Isaacson:1968hbi,
    author = "Isaacson, Richard A.",
    title = "{Gravitational Radiation in the Limit of High Frequency. I. The Linear Approximation and Geometrical Optics}",
    doi = "10.1103/PhysRev.166.1263",
    journal = "Phys. Rev.",
    volume = "166",
    pages = "1263--1271",
    year = "1968"
}

@article{Abramo:1997hu,
    author = "Abramo, L. Raul W. and Brandenberger, Robert H. and Mukhanov, Viatcheslav F.",
    title = "{The Energy - momentum tensor for cosmological perturbations}",
    eprint = "gr-qc/9704037",
    archivePrefix = "arXiv",
    reportNumber = "BROWN-HET-1046",
    doi = "10.1103/PhysRevD.56.3248",
    journal = "Phys. Rev. D",
    volume = "56",
    pages = "3248--3257",
    year = "1997"
}

@article{Ning:2023ybc,
    author = "Ning, Sirui and Sou, Chon Man and Wang, Yi",
    title = "{On the decoherence of primordial gravitons}",
    eprint = "2305.08071",
    archivePrefix = "arXiv",
    primaryClass = "hep-th",
    doi = "10.1007/JHEP06(2023)101",
    journal = "JHEP",
    volume = "06",
    pages = "101",
    year = "2023"
}

@article{Tsamis:1996qq,
    author = "Tsamis, N. C. and Woodard, R. P.",
    title = "{Quantum gravity slows inflation}",
    eprint = "hep-ph/9602315",
    archivePrefix = "arXiv",
    reportNumber = "CPTH-S372-0995, CRETE-95-11, UFIFT-HEP-95-17",
    doi = "10.1016/0550-3213(96)00246-5",
    journal = "Nucl. Phys. B",
    volume = "474",
    pages = "235--248",
    year = "1996"
}

\appendix

\section{Time-dependent computation}
\label{sec: Time-dependent computation}

The equation of motion for $\phi_2$ can be found in Ref. \cite{Bertacca:2024zfb} and describes the dynamics of the gravitational potential on all scales. In Refs. \cite{Traforetti:2025cax,Abdelaziz:2026qyk}, we developed the calculation for the two- and three-point functions, taking the super-horizon scales solution in which $\phi_2$ is constant over time. 
In this Appendix, we provide the ingredients to obtain the all-scale solution of $\phi_2$, going beyond the very-large-scale limit used in \cite{Traforetti:2025cax,Abdelaziz:2026qyk}.

Starting from Eq. (9) of \cite{Bertacca:2024zfb}, the fluid closure used in the present calculation is deliberately restricted: we set to zero $\Pi_2=\delta^2S=0$ and the adiabatic speed of sound is set to $c_s^2=1$. 
The source term on the RHS is quadratic in the first-order tensor perturbation $\chi_{1ij}$ and, using $\mathcal H=-1/\eta$, it reads 
\begin{widetext}
\begin{align}
S(\eta)=&-\frac{9}{\eta^2}\nabla^{-4}\partial_i\partial^j \mathcal A^i_j+\frac{3}{2}\frac{1}{\eta}\nabla^{-4}\partial_i\partial^j {\mathcal A^i_j}' 
-\frac{1}{2}\nabla^{-2}\partial_i\partial^j\mathcal A^i_j \nonumber \\  
&-\frac{6}{\eta^2}\nabla^{-2}\Bigg[-\frac{3}{4}\chi_1^{lk,m}\chi_{1kl,m}
- \frac{1}{2} \chi_1^{kl}\nabla^2 \chi_{1lk} 
+ \frac{1}{2}\chi_{1,l}^{km}\chi_{1m,k}^l+ \frac{1}{2}\chi_{1km}'   {\chi_{1km}'} 
- \frac{3}{2}\nabla^{-2}\partial_i\partial_j \left({\chi_{1k}^{i}}'   {\chi_{1kj}'}\right)\Bigg] \nonumber\\  
&+\frac{3}{8}\chi_{1,k}^{ml}\chi_{1lm}^{,k}
-\frac{3}{8}{\chi_1^{kl}}' {\chi_{1kl}'}
-\frac{1}{4}\chi_{1k,l}^m\chi_{1,m}^{lk} \nonumber\\
&+\frac{1}{\eta}\nabla^{-2}\Bigg[-\frac{3}{2}{\chi_1^{lk,m}}'\chi_{1kl,m} 
+ \frac{1}{2}{\chi_1^{kl}}' \nabla^2 \chi_{1lk} 
- \frac{1}{2}\chi_1^{kl}\nabla^2{\chi_{1lk}}' + \frac{1}{2}{\chi_{1,l}^{km}}'\chi_{1m,k}^l + \frac{1}{2}\chi_{1,l}^{km}{\chi_{1m,k}^l}'
 +2\frac{1}{\eta}{\chi_{1km}}' {\chi_{1km}'}\Bigg] 
\nonumber\\
&
-3\frac{1}{\eta}\nabla^{-4} \left(2\frac{1}{\eta}{\chi_{1k,j}^{i}}'{\chi_{1k,i}^j}'
+{\chi_{1k,i}^j}' \nabla^2 \chi_{1k,j}^{i}\right) +\frac{1}{2}\nabla^{-2}\partial_i\partial_j\left({\chi_{1k}^{i}}'   {\chi_{1kj}'}\right) \nonumber\\
&+\Bigg[-\frac{1}{8}{\chi_1^{ik}}'\chi_{1ki}'
+\frac{1}{\eta}\chi_1^{ik}\chi_{1ki}'
+\frac{1}{2}\chi^{mk}_1\nabla^2\chi_{1mk}
+\frac{3}{8}\chi^{ml}_{1,k}\chi_{1lm}^{,k}
-\frac{1}{4}\chi_{1k,l}^m\chi_{1,m}^{lk} \Bigg]\,,
\label{eq:source}
\end{align}
where $\mathcal A^i_j$ is defined by
\begin{align}
\mathcal A^i_j=&\tfrac12\chi_1^{lk,i}\chi_{1kl,j}+\chi_1^{kl}\chi_{1lk,j}^{,i} -\chi_1^{kl}\chi_{1l,jk}^i-\chi_1^{kl}\chi_{1lj,k}^{,i}\quad+\chi_1^{kl}\chi_{1j,kl}^i+\chi_{1,l}^{ik}\chi_{1jk}^{,l}
-\chi_{1,l}^{ki}\chi_{1j,k}^l\,.
\end{align}
\end{widetext}
In Fourier space, the equation of motion reads
\begin{equation}
\phi_{2k}''-\frac{6}{\eta}\,\phi_{2k}'+\Big(\frac{6}{\eta^2}+k^2\Big)\phi_{2k}=S_k(\eta).
\label{eq:phi2}
\end{equation}
Solving the homogeneous equation, it can be shown that the Green's function takes the following expression
\begin{equation}
G_k(\eta,\tilde\eta)=\frac{k\eta^4}{\tilde\eta^2}\big[j_2(k\tilde\eta)\,y_2(k\eta)
-j_2(k\eta)\,y_2(k\tilde\eta)\big]\,,
\label{eq:green}
\end{equation}
where $j_2(x)$ and $j_2(x)$ are the spherical Bessel functions of the second kind. We expand the first-order tensors using the usual helicity decomposition
\begin{align}
\chi_{1ij}(\mathbf{x},\eta)=&\sum_{\lambda=\pm}\int\frac{d^3q}{(2\pi)^3}e^{i\mathbf{q}\cdot\mathbf{x}}e_{ij}^{\lambda}(\hat{\mathbf q})\chi_\lambda(\mathbf q)T_q(\eta),
\label{eq: expansion chi}
\end{align}
where $e^\lambda_{ij}(\mathbf{q})$ are the polarization tensors, with $\lambda$ is the polarization index, $\chi_\lambda(\mathbf q)$ denotes the initial mode amplitude and $T_q(\eta)$ includes its time-dependence. For tensor perturbations in a dS background, the Bunch-Davies solution in dS can be written as $T_q(\eta)=H(\eta_q)(1+iq\eta)e^{-iq\eta}$ and the mode freezes outside the horizon. The overall amplitude, including its dependence on the energy scale of the dS, is encoded in $\chi_\lambda(\mathbf q)$. 

Substituting \eqref{eq: expansion chi} into the quadratic source in Eq.\eqref{eq:source}, its Fourier transform can be expressed as
\begin{align}
    S_k(\eta)=\sum_{\lambda_1,\lambda_2}\int\dfrac{d^3\mathbf{k}_1}{(2\pi)^3}\chi_{\lambda_1}(\mathbf{k}_1)\chi_{\lambda_2}(\mathbf{k}_2)f_{\lambda_1\lambda_2}(\mathbf{k}_1,\mathbf{k}_2;\eta)\,,
    \label{eq: source in Fourier schematic}
\end{align}
where the momenta form a triangle in momentum space, so they satisfy $\mathbf{k}_1+\mathbf{k}_2=\mathbf{k}$. 
The source function $f$ contains the dependence on the polarization and the geometry, together with the time-dependence inherited from $T_q(\eta)$. Schematically, we can write
\begin{align}
f_{\lambda_1\lambda_2}(\mathbf{k},\mathbf{k}_1;\eta)=&\sum_{a,b=0}^{1}C^{\lambda_1\lambda_2}_{(a,b)}(\mathbf{k},\mathbf{k}_1;\eta)\nonumber\\
&\times T_{k_1}^{(a)}(\eta)T_{|\mathbf{k}-\mathbf{k}_1|}^{(b)}(\eta),
\label{eq:pairwise source kernel}
\end{align}
where $T_q^{(0)}(\eta)\equiv T_q(\eta)$ and $T_q^{(1)}(\eta)\equiv dT_q(\eta)/d\eta$, since the source \eqref{eq:source} contains at most one conformal-time derivative acting on $\chi_{1ij}$. The coefficients $C$ in \eqref{eq:pairwise source kernel} collect the momentum-dependent factors and the explicit conformal-time dependence in Eq. \eqref{eq:source}. 
If we collectively classify the possible helicity pairs with $\sigma=s_{\lambda_1}s_{\lambda_2}=\pm1$, we have that each class contains two equivalent configurations. Then, we can write the kernel as
\begin{equation}
I_\sigma(\mathbf{k},\mathbf{k}_1;\eta_\ast)=\int_{\eta_i}^{\eta_\ast}d\tilde\eta\;w(\tilde\eta)\,
G_k(\eta_\ast,\tilde\eta)\,f_\sigma(\mathbf{k},\mathbf{k}_1;\tilde\eta)\,,
\label{eq: kernel}
\end{equation}
which is the time-integrated response of $\phi_2$ to the graviton pair evaluated at $\eta_\ast$ and it includes the Green's function, the source structure and the polarization contractions. In the expression above, $w(\eta)$ is the smooth switch-on weight described below and we assumed $\phi_{2k}=\phi_{2k}'=0$ at $\eta_i$. 

Using $\langle\phi(\mathbf{k})\phi(\mathbf{k}')\rangle=\frac14\langle\phi_2(\mathbf{k})\phi_2(\mathbf{k}')\rangle$ and performing the Wick contractions of the first-order tensor modes, the dimensionless spectrum evaluated at $\eta_\ast$ can be written as
\begin{align}
\Delta_\phi^2(k,\eta_\ast)\propto  k^3\int d^3\mathbf{k}_1 & \sum_{\sigma=\pm}2\,\left|I_\sigma(\mathbf{k},\mathbf{k}_1;\eta_\ast)\right|^2\nonumber\\
&\times\mathcal{P}_h(k_1)\mathcal{P}_h(|\mathbf{k}-\mathbf{k}_1|)\,, 
\label{eq: dimensionless spectrum of scalars}
\end{align}
where the factor 2 accounts for the two equivalent configurations in each class $\sigma=\pm$ and the other momentum-independent factors are absorbed into the overall normalization. The tensor power spectra have been taken out, and they read $\mathcal{P}_h(k_1)=k_1^{-3}\mathcal{A}_h(k_\ast)(k_1/k_\ast)^{n_T}$, with $\mathcal{A}_h$ the tensor amplitude at the pivot and $n_T=0$ in dS. Since the three momenta form a triangle in momentum space, they satisfy $|k_1-k_2|\leq k\leq k_1+k_2$. The loop integral is restricted to $k_1,k_2\geq k_{\rm min}$ and $k_1\leq X_{\rm uv}k$, where $X_{\rm uv}$ is the dimensionless UV-cutoff kept fixed during the sweep over the external scale $k$. The infrared cutoff, instead, represents the first comoving mode available from the onset of inflation and is therefore held fixed as $k$ is varied. Its dimensionless form is $\mu(k)\equiv k_{\rm min}/k$. Every ingredient scales homogeneously under $k_i\rightarrow\lambda k_i$ and $\eta\rightarrow\eta/\lambda$. Hence, setting $\lambda=1/k$ so $x_c=-\eta_\ast$, the dimensionless power spectrum can be written schematically as
\begin{align}
\Delta_\phi^2(k,\eta_\ast)= \mathcal{A}^2_h(k_\ast)\left(\dfrac{k}{k_\ast}\right)^{2n_T} \mathcal{I}\left(\mu;X_{\rm uv},x_c,w,c_s\right)\,,
\label{eq: final expression scalar spectrum}
\end{align}
where $\mathcal{I}$ denotes the results of the 2D integration of the squared kernel and the time integral, and the semicolon is meant to emphasize that, within a given numerical run, $X_{\rm uv},x_c,$ and the background parameters are held fixed. It is clear from Eq. \eqref{eq: final expression scalar spectrum} that the cutoffs $k_{\rm min}$ and $k_{\rm max}$ do not play equivalent roles in the numerical prescription. In fact, the only dilation-breaking scale is the comoving infrared cutoff. This explains the expression of the spectral slope \eqref{eq: spectral slope} given in the main text. 

In Section \ref{sec:Power spectrum in quasi de Sitter}, we introduce $\eps$ parametrizing the drift in the Hubble parameter $H$; therefore, each tensor leg is additionally weighted by $(q/k_\ast)^{-\eps}$. Since in the IWI framework the scalar is a graviton bilinear, the four factors entering the scalar power spectrum yield an overall contribution $(k/k_\ast)^{-4\eps}$. 

In Eq. \eqref{eq: kernel}, we start the time integration at $\eta_i=-Rx_c/k$, with $R=20$ which determines how far back in time $\eta_i$ is compared to the final evaluation time $\eta_\ast$. To suppress spurious boundary contributions from the rapidly oscillating sub-horizon source, we smoothly switch-on the source using a smooth switch-on weight $w(\tilde\eta)=\sin^2\left[\pi k(\tilde\eta-\eta_i)/2\ell\right]$ for $0\leq k(\tilde\eta-\eta_i)<\ell$, and $w(\tilde\eta)=1$ for $k(\tilde\eta-\eta_i)\geq\ell$, with $\ell=2$. 
We test the sensitivity of the result to $R$ and $\ell$, finding weak dependence on the initial time and switch-on prescription.
The time integral of Eq. \eqref{eq: kernel} is evaluated on a uniform grid with $\Delta\eta=0.2/[(2X_{\rm uv}+1)k]$, while the momentum integrals use a combination of logarithmic and linear grids. 
We have also tested the numerical convergence of the calculation by refining the momentum grid and the time discretization, finding that the dependence is weak for $\mathcal{I}(\mu;X_{\rm uv},x_c,w,c_s)$.

In the numerical evaluation, e.g., using the baseline with $k_{\rm min}=0.01\;\rm Mpc^{-1}$, $x_c=1$ and $X_{\rm uv}=25$, we find that after horizon crossing, the potential $\phi_2$ approaches a constant, consistent with the large-scale solution discussed in Ref. \cite{Traforetti:2025cax}. The cut-off dependent shape function $\mathcal{I}(\mu;X_{\rm uv},x_c,w,c_s)$ in \eqref{eq: final expression scalar spectrum} only introduces a tiny blue tilt contribution (see the main text) \footnote{We also performed reading time, $x_c$, and UV truncation, $X_{\rm uv}$, tests in order to check the robustness of our choices. In particular, the growth of $\mathcal{I}$ as roughly $X_{\rm uv}^3$ shows that its normalization is set by the loop momenta that are still sub-horizon at the crossing, i.e., the vacuum dressing discussed in Section \ref{sec:smallscales}.}.

\section{The graviton effective fluid}
\label{sec: The graviton effective fluid}

This Appendix collects the derivation behind Eqs.~\eqref{eq:fluid} and \eqref{eq:epsself}, starting from the second-order gravitational-wave effective energy-momentum tensor of Abramo, Brandenberger and Mukhanov (ABM) \cite{Abramo:1997hu}. In the absence of matter fluctuations, ABM's Eq.~(82) gives the effective energy density
\begin{equation}
8\pi G\,\rho_{\rm GW}=H\langle\dot\chi_{ij}\chi_{ij}\rangle +\tfrac18\big[\langle\dot\chi_{ij}\dot\chi_{ij}\rangle +a^{-2}\langle\chi_{ij,k}\chi_{ij,k}\rangle\big]\,,
\label{eq:abmrho}
\end{equation}
with dots denoting cosmic-time derivatives. Their effective pressure [ABM Eq.~(88)] contains in general a background term proportional to $\rho^{(0)}+p^{(0)}$. Since in dS the background Friedmann equation gives $\rho^{(0)}+p^{(0)}=0$, the identification $p_{\rm GW}=-\tfrac13\tau^i{}_i$ is exact, and isotropic averaging of ABM Eq.~(83) yields [their Eq.~(89)]
\begin{equation}
8\pi G\,p_{\rm GW}=\tfrac{7}{24}a^{-2}\langle\chi_{ij,k}\chi_{ij,k}\rangle -\tfrac{5}{24}\langle\dot\chi_{ij}\dot\chi_{ij}\rangle \,.
\label{eq:abmp}
\end{equation}

\begin{figure}[t]
\centering\includegraphics[width=\columnwidth]{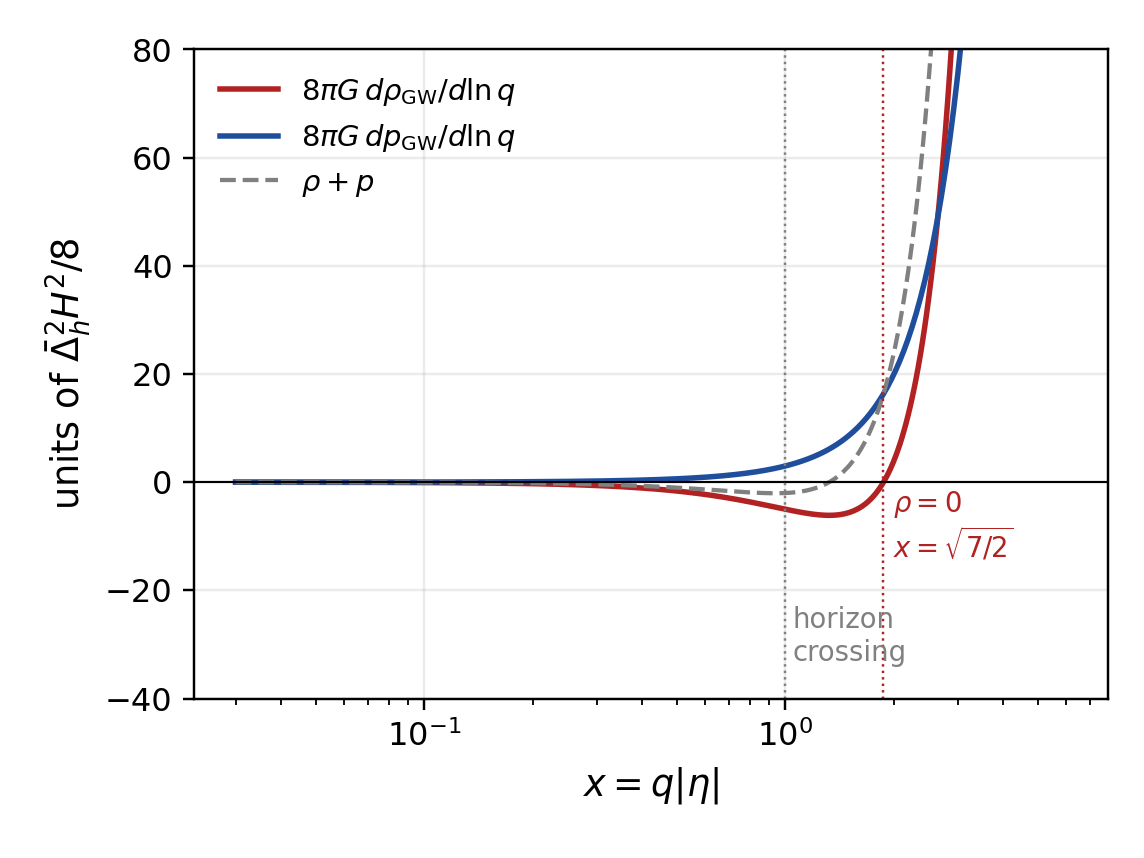}
\caption{The graviton effective fluid per $\ln q$. We plot Eq.~\eqref{eq:fluid} in units of
$\Dh H^2/8$ as a function of $x\equiv q|\eta|$: the energy density profile $d\rho_{\rm GW}/d\ln q$ as a red line, the pressure profile $dp_{\rm GW}/d\ln q$ as a blue line and their sum as a dashed line. The graviton population behaves as radiation-like on sub-horizon scales ($x\gg1$), negative energy with
$w\to-1/3$ on super-horizon scales ($x\ll1$). The density changes sign at $x=\sqrt{7/2}$.}
\label{fig:backreaction}
\end{figure}

We consider the Bunch-Davies mode $h_q(\eta)=h_q^{\rm fr}(1+iq\eta)e^{-iq\eta}$, where $h_q^{\rm fr}$ is the frozen amplitude. With $x=q|\eta|$ and the frozen tensor power per logarithmic interval defined as $\Dh\equiv d\langle\chi^{\rm fr}_{ij}\chi^{\rm fr}_{ij}\rangle/d\ln q$, the three correlators per $\ln q$ appearing in Eqs. \eqref{eq:abmrho} and \eqref{eq:abmp} evaluate to
\begin{align}
\frac{d\langle\chi_{ij}\chi_{ij}\rangle}{d\ln q}&=\Dh(1+x^2)\,,\nonumber\\
\frac{d\langle\dot\chi_{ij}\dot\chi_{ij}\rangle}{d\ln q}&=H^2\Dh\,x^4\,,\\
\frac{d\langle\dot\chi_{ij}\chi_{ij}\rangle}{d\ln q}&=-H\Dh\,x^2 \,,\nonumber
\end{align}
where we used $\dot h_q=h_q'/a$, $h_q'=h_q^{\rm fr}q^2\eta\,e^{-iq\eta}$ and $q^2/a^2=H^2x^2$. Substituting into Eqs.~\eqref{eq:abmrho} and \eqref{eq:abmp}, we obtain the per-mode profiles
\begin{align}
8\pi G\,\frac{d\rho_{\rm GW}}{d\ln q}&=\frac{\Dh H^2}{8}\,x^2(2x^2-7)\,,\nonumber\\
8\pi G\,\frac{dp_{\rm GW}}{d\ln q}&=\frac{\Dh H^2}{24}\,x^2(2x^2+7)\,,
\label{eq:fluid appendix}
\end{align}
which reproduces Eq.~\eqref{eq:fluid} of the main text, and that the direct symbolic expansion of $G^{(2)}$ in Sec.~\ref{sec:classical channel} also gives. Checking the limiting behavior of Eq. \eqref{eq:fluid appendix}, we see that for $x\gg1$, $p/\rho\to1/3$ (the Isaacson radiation fluid \cite{Isaacson:1968hbi}), while for $x\ll1$ the density is negative with $p/\rho\to-1/3$. We show the behavior of the graviton population in Figure \ref{fig:backreaction}.

The next step is to determine how strongly this effective fluid changes the background Hubble rate. The background acceleration equation is
\begin{equation}
    \dot H=-4\pi G(\rho+p)=-\frac{\rho+p}{2m_{\rm pl,red}^2}\,.
\end{equation}
The contribution generated by the tensor population is therefore
\begin{equation}
\eps_{\rm self}\equiv-\frac{\dot H_{\rm GW}}{H^2}=\frac{\rho_{\rm GW}+p_{\rm GW}}{2m_{\rm pl,red}^2H^2}\,.
    \label{rn:epsilonselfdef}
\end{equation}
From Eq. \eqref{eq:fluid appendix}, we then obtain
\begin{equation}
\frac{d\eps_{\rm self}}{d\ln q}=\frac{\Dh}{24}\,x^2(4x^2-7)\,.
\label{eq: eps self appendix}
\end{equation}
Integrating Eq. \eqref{eq: eps self appendix} at fixed time from the infrared up to a matching scale $X$ gives
\begin{equation}
\eps_{\rm self}(X)=\frac{\Dh}{48}\big(2X^4-7X^2\big).
\label{eq:epsselfX}
\end{equation}
At $X=1$ this is $\eps_{\rm self}(1)=-\tfrac{5}{48}\Dh$, while over $X\in[1,2]$ it spans $-0.128\lesssim \eps_{\rm self}/\bar\Delta_h^2\lesssim0.083$, with the minimum at $X=\sqrt7/2$ and crossing zero at $X=\sqrt{7/2}\simeq1.87$.
Hence, we can conclude that the overall magnitude of the graviton backreaction is $|\eps_{\rm self}|\lesssim0.13\,\Dh$.

\end{document}